\documentstyle[epsfig, twocolumn]{iso98}

\newcommand{\thepapername}{}

\begin{document}

\setlength{\parindent}{0pt}
\setlength{\parskip}{ 10pt plus 1pt minus 1pt}
\setlength{\hoffset}{-1.5truecm}
\setlength{\textwidth}{ 17.1truecm }
\setlength{\columnsep}{1truecm }
\setlength{\columnseprule}{0pt}
\setlength{\headheight}{12pt}
\setlength{\headsep}{20pt}
\pagestyle{veniceheadings}

\title{\bf SPATIAL DISTRIBUTION OF UNIDENTIFIED INFRARED BANDS AND 
EXTENDED RED EMISSION IN THE COMPACT GALACTIC HII REGION SH~152 
\thanks{ISO is an ESA
project with instruments funded by ESA Member States (especially the PI
countries: France, Germany, the Netherlands and the United Kingdom) and
with the participation of ISAS and NASA.}
\thanks{Partly based on observations made at Observatoire de Haute Provence du
CNRS}}

\author{{\bf S.~Darbon$^1$, A.~Zavagno$^2$, C.~Savine$^1$, V.~Ducci$^2$, 
J.-M.~Perrin$^1$, J.-P.~Sivan$^1$} \vspace{2mm} \\
$^1$Observatoire de Haute Provence du CNRS, 04870 Saint-Michel l'Observatoire, 
France. \\
$^2$Observatoire de Marseille, 2 Place Le Verrier, 13248 Marseille Cedex 4, 
France. }

\maketitle

\begin{abstract}

We present visible and near IR images of the compact H\,{\sc{ii}} region Sh~152.
Some of these images reveal the presence of Extended Red Emission (ERE) around
698~nm and emission from Unidentified Infra Red Bands (UIRBs) at 3.3 and
6.2\,$\mu$m. Other images show the near infrared (7-12\,$\mu$m) continuous
emission of the nebula. The ERE emission is found to coincide with the ionized
region and significantly differ from the UIRBs location. Also some evidence
is found in favor of grains as carriers for ERE.
\vspace {5pt} \\


 Key~words: H\,{\sc{ii}} regions, ERE, UIRBs 

\end{abstract}

\section{INTRODUCTION}

Extended red emission (ERE) is a continuous emission band observed in the 
red part (600--800~nm) of the spectrum of various astrophysical objects 
like reflection nebulae (\cite{sch80}, \cite{wb90}), planetary nebulae 
(\cite{fw92}), H\,{\sc{ii}} regions (\cite{sp93}, \cite{dps98}), high-latitude 
galactic cirrus clouds (\cite{sg98}), the halo of the galaxy M82 (\cite{pds95})
 and also in the diffuse galactic interstellar medium (\cite{gwf98}). This 
emission can be attributed either to Hydrogenated Amorphous Carbon (HAC) 
grains (\cite{wat82}, \cite{fw93}) or silicon grains (\cite{wgf98}, 
\cite{led98}). 
A series of emission bands in the 3--16\,$\mu$m range, the so-called 
UIRBs, is also observed in dusty environments and commonly attributed 
to Polycyclic Aromatic Molecules (PAH)(\cite{pl89}, \cite{all89}) and/or 
carbonaceous materials (\cite{pap89}). 
In particular, the existence (or absence) of a spatial correlation
between IURBs and ERE might be useful to put constrains on the nature of the
carriers. Compact H\,{\sc{ii}} regions are bright and dusty objects well 
suited for this kind of study. \\
This is the reason why we have carried out an observational program for imaging
compact H\,{\sc{ii}} regions at visible and infrared wavelengths in order to
detect and to map respectively ERE and UIRBs. This paper reports on the results
obtained for Sh~152.

\section{OBSERVATIONS AND DATA REDUCTION}

Infrared images of Sh~152 were obtained with \mbox{ISOCAM} in june 1997, during 
ISO revolution 563. These include UIRBs images at 3.3 and 6.2\,$\mu$m and
four continuum images taken with the ISOCAM circular variable filter (CVF) at
6.911, 8.222, 10.52 and 12.00\,{$\mu$m}. These observations and data reduction
are described in \cite{zd98}. In particular, the 3.3 and
6.2\,$\mu$m images presented in this paper were corrected for the adjacent
continuum emission.\\
Visible images in 
the 500--850~nm range were obtained, in october 1997, with a 1024 $\times$ 
1024 thinned back-illuminated Tektronix CCD camera mounted at the Newton focus 
of the 120~cm telescope of the Observatoire de Haute Provence. 
Four interference filters with a FWHM~$\simeq$10~nm centered on 528.2, 
612.0, 697.5 and 812.5~nm were used. These filters were chosen to
isolate the continuum emission of Sh~152 and to avoid nebular and night sky 
emission lines.
For each continuum filter, twenty-five 15 min exposure time frames were 
obtained and co-added, yielding a resulting image of six hours exposure time. 
Standard data reduction was performed using ESO-MIDAS software. It includes : 
dark current subtraction, flat fielding, airmass and interstellar extinction 
corrections, deconvolution by point spread function. According to spectroscopic 
observations of ERE in H\,{\sc{ii}} regions (\cite{sp93}), the emission excess 
in the 697.5 and/or 812.5~nm filters should be attributed to ERE. Actually the 
best contrasted results were obtained by making the difference between the 
697.5 and 612.0~nm images. The resulting image was considered as giving the 
spatial distribution of ERE over Sh~152.
\newpage
\begin{figure}[!ht]
  \begin{center}
    \leavevmode
  \centerline{\epsfig{file=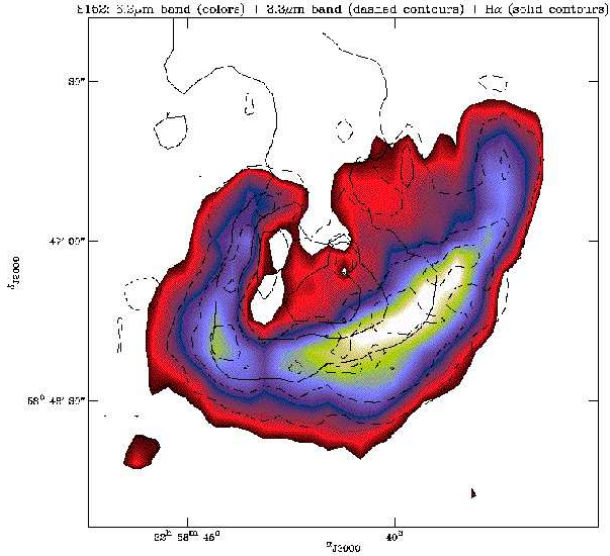,width=8cm}}
  \end{center}
  \caption{\em 6.2\,$\mu$m (colors), 3.3\,$\mu$m emission bands (dashed 
    contours) and H$\alpha$ emission (solid contours) in Sh~152}
  \label{fig:fig1}
\end{figure}
\begin{figure}[!ht]
  \begin{center}
    \leavevmode
   \centerline{\epsfig{file=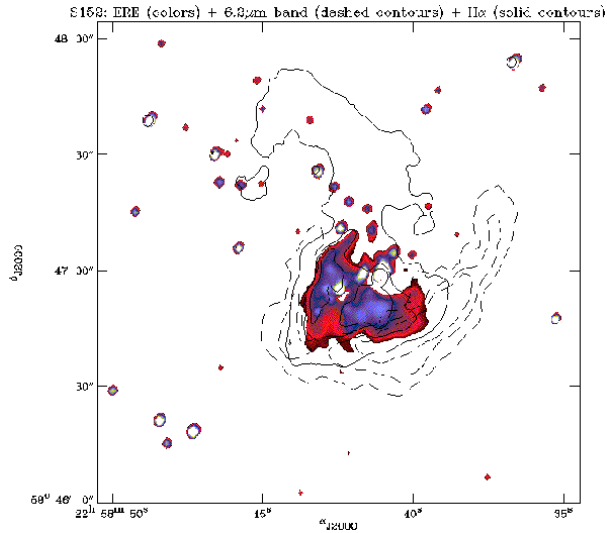,width=8cm}}
  \end{center}
  \caption{\em ERE distribution (colors), 6.2\,$\mu$m emission band (dashed 
    contours) and H$\alpha$ emission (solid contours) in Sh~152}
  \label{fig:fig2}
\end{figure}
\begin{figure}[!ht]
  \begin{center}
    \leavevmode
  \centerline{\epsfig{file=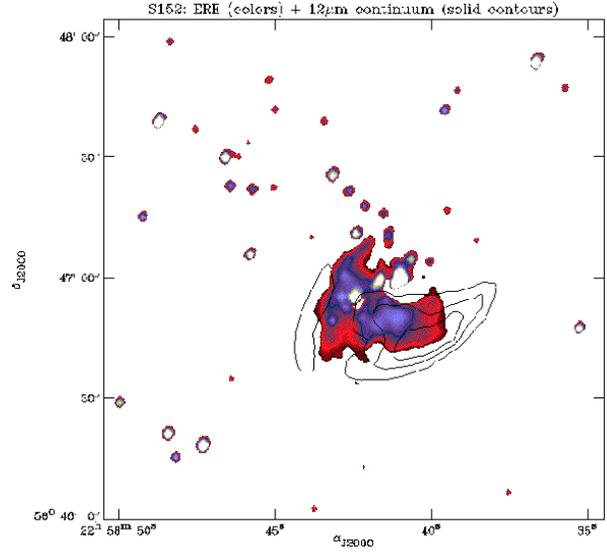,width=8cm}}
  \end{center}
  \caption{\em ERE distribution (colors) in Sh~152 and 12\,$\mu$m  continuum 
    image (solid contours)}
  \label{fig:fig3}
\end{figure}
\begin{figure}[!ht]
  \begin{center}
    \leavevmode
  \centerline{\epsfig{file=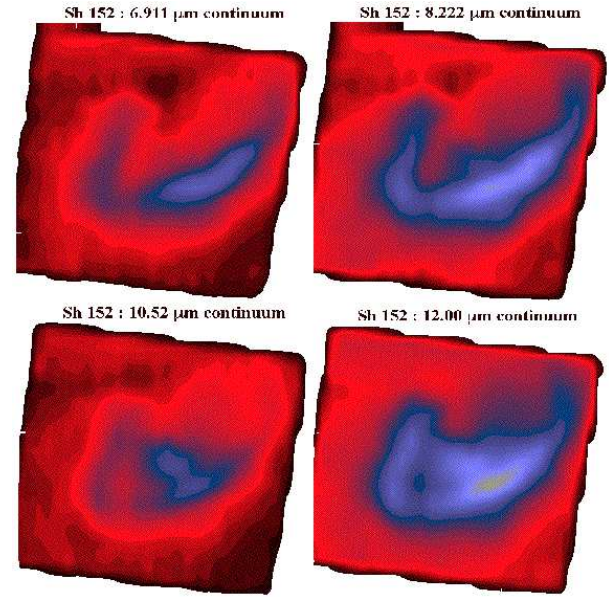,width=8cm}}
  \end{center}
  \caption{\em Continuum emission in Sh~152 observed with the ISOCAM CVF at 
  6.911, 8.222, 10.52 and 12.00\,$\mu$m}
  \label{fig:fig4}
\end{figure}
\section{RESULTS AND DISCUSSION}
 The 3.3\,$\mu$m, 6.2\,$\mu$m and H$\alpha$ images of Sh~152 are displayed in
Figure~\ref{fig:fig1}. It can be seen that the 3.3 and 6.2\,$\mu$m emission
bands have the same spatial distribution over the entire nebula. This suggests
their carriers could be the same. Also, it appears from Figure~\ref{fig:fig1}
that the infrared emissions arise from regions located outside the ionized
regions (traced by H$\alpha$ emission).

Figure~\ref{fig:fig2} presents the spatial distribution of ERE superimposed on 
the 6.2\,$\mu$m band image and the H$\alpha$ image. 
ERE is found to coincide with the H$\alpha$ emission but significantly differs  
from that of the 6.2\,$\mu$m emission band. Hydrogen environment and UV 
radiation are well suited to induce luminescence from HAC grains (\cite{fw93}). 

Figure~\ref{fig:fig3} presents the spatial distribution of ERE 
superimposed on a 12\,$\mu$m continuum image. It can be seen that the 
12\,$\mu$m emission extends over the area where the ERE intensity reaches its
maximum. This coincidence is in favor of grains as carriers of the ERE
because (i) the 12\,$\mu$m emission is thought to be the short wavelength part
of a strong thermal emission from cold grains (see, for example, IR spectra of 
ultra compact
galactic H\,{\sc{ii}} regions presented by \cite{roe98})and (ii) because such 
cold grains can exist in Sh~152 at the distance from the 
exciting star where the observed coincidence occurs.\\  
In effect, according to \cite{lp97}, the temperature of a dust solid particle located at a
distance 10$^4$~R$_*$~= 2.10$^2$~AU = 10$^{-3}$~pc from an O9.5 V star of 
radius R$_* \simeq 2.10^{-2}$~AU, would be of 200K for a silicate grain or 
400K for a carbonaceous grain. The region in Sh~152 where ERE maximum and 
12~$\mu$m emission coincide is in fact much
farther from the star (about 0.2 pc, assuming a distance of 3.5~kpc for 
Sh~152 (\cite{ht81})) than in the calculations so that, although
the exciting star of Sh~152 is slightly hotter than an O9.5V star 
(\cite{hm90}), we can assume that cold grains do exist in the area.
 
Figure~\ref{fig:fig4} presents the four continuum images of Sh~152 at 6.911, 
8.222, 10.52 and 12.00\,$\mu$m taken with ISOCAM. In these images, the flux 
is normalized to the maximum observed in the LW6 filter, centered at 
7.7\,$\mu$m. At the location of the ERE maximum, the infrared images show flux 
values increasing with wavelength : this is in agreement with thermal emission 
from cold grains (note that the 10.52\,$\mu$m flux might be contaminated by the 
[SIV] 10.54\,$\mu$m emission line).

\section{CONCLUSION}
Visible and infrared images of Sh~152 allowed us to compare the spatial 
distribution of ERE and UIRBs and to deduce basic physical properties. We
found that :\\
\begin{itemize}
\item the spatial distribution of the two UIRBs at 3.3 and 6.2\,$\mu$m are the 
same, suggesting similar properties for their carriers
\item the UIRBs emission peak is located at the border of the ionized region
\item the ERE emission coincides with the ionized region and significantly 
differs from the UIRBs location. 
\item the continuum emission observed between 10.5 and 12\,$\mu$m is 
coincident with ERE emission and possibly due to cold grains, which is in 
favor of grains as carriers of the ERE. 
\end{itemize}   
Nevertheless, the exact nature of the ERE carriers will only be constrained 
using spectroscopic data and comparing the ERE band shape with laboratory 
experimental data (Darbon et al., in preparation) 
 


\end{document}